\documentclass[11pt]{article}
\usepackage[a4paper,margin=28mm]{geometry}
\usepackage{fontspec}
\usepackage{amsmath,amssymb,amsthm,mathtools}
\usepackage{microtype}
\usepackage{mathrsfs}
\usepackage{enumitem}
\usepackage[colorlinks=true,linkcolor=blue!45!black]{hyperref}
\usepackage{xcolor}
\newcommand{\R}{\mathbb R}
\newcommand{\su}{\mathfrak{su}(2)}
\newcommand{\F}{\mathcal F}
\newtheorem{lemma}{Lemma}[section]
\newtheorem{theorem}[lemma]{Theorem}
\newtheorem{proposition}[lemma]{Proposition}
\theoremstyle{definition}
\newtheorem{definition}[lemma]{Definition}
\numberwithin{equation}{section}
\title{Bolognesi's Conjecture for Nonnegative-Charge Iterated Magnetic Bags}
\author{Weifeng Sun \and Wenhan Wang}
\date{}
\begin{document}
\maketitle

\begin{abstract}
We establish a magnetic-bag limit for smooth $SU(2)$ Bogomolny monopoles
associated with finitely many nested smooth surfaces in $\R^3$.
This proves Bolognesi's magnetic-bag conjecture for iterated magnetic bags
with nonnegative charge density on each surface.
\end{abstract}

\section{Introduction}\label{sec:introduction}

Bolognesi's \emph{magnetic-bag conjecture} arose from his study of
$SU(2)$ monopoles with large topological charge~\cite{Bolognesi2006}.
Harland, Palmer and S\"amann~\cite{HarlandPalmerSaemann2012} formulated the conjecture as follows:
A prescribed magnetic bag should arise as a limit of smooth $SU(2)$ monopoles
as the charge and coupling grow proportionally.
Manton~\cite{Manton2012} described spherical multilayer bags in the abelian approximation.
Taubes constructed smooth solutions with magnetic-bag-like behavior
when the bag is bounded by a single sphere~\cite{Taubes2013}.

In this paper, we construct smooth $SU(2)$ monopoles that converge to any nonnegative-charge
iterated magnetic bag as the charge tends to infinity.
This proves Bolognesi's magnetic-bag conjecture for every nonnegative-charge
iterated magnetic bag.

Our approach is inspired by Taubes' work but uses a different strategy to establish convergence.
Taubes~\cite{Taubes2013} bounded the thickness of the spherical shell containing
the Higgs zeros in terms of the monopole charge. We instead fix a strictly positive
approximation to both the surface densities and the limiting Higgs
function, construct monopoles close to the approximation, and then recover
the desired magnetic bag by a diagonal argument.
This substantially simplifies the proof: Correcting an approximate Bogomolny
solution to a genuine one requires only a single application of the Banach
inverse function theorem.
The cost, compared to Taubes' iterative scheme, is that we do not obtain
an explicit convergence rate.\footnote{%
Strictly speaking, one can also try to track quantitative convergence
rates in our construction. GPT-6 Astra
was tasked with this and reported the following estimates: For strictly
positive surface densities and limiting Higgs function,
the Higgs norms \(|\Phi|\) and projected curvatures \(\F\)
converge at rate \(O(N^{-1/2})\) in each fixed \(C^r\) norm on compact
sets away from the surfaces. For general nonnegative data, it reported
an \(O(N^{-1/40})\) rate of convergence. Every Higgs zero lies
within \(o(N^{-1})\) of the prescribed surfaces.
Since these quantitative refinements are not essential/interesting/difficult,
the authors choose not to verify
these AI-generated estimates and do not present them as established
conclusions. Readers interested in these explicit estimates of convergence rate, which may not exist,
should establish them themselves as an exercise.}

We now set the stage.

Let $A$ be an $\su$-valued one-form and $\Phi$ an adjoint Higgs field.
For a coupling constant $e>0$, the Bogomolny equation is written as
\[
 F_{A,e}=*d_{A,e}\Phi,\qquad
 F_{A,e}=dA+\frac e2[A\wedge A],\qquad
 d_{A,e}\Phi=d\Phi+e[A,\Phi],
\]
where $*$ is the Euclidean Hodge star.
A configuration $(A,\Phi)$ satisfying this equation is called a \emph{monopole}.
The equation is invariant under smooth $SU(2)$ gauge transformations
$g:\R^3\to SU(2)$, acting by
\[
 A^g=g^{-1}Ag+e^{-1}g^{-1}dg,\qquad \Phi^g=g^{-1}\Phi g.
\]

Fix $n\geq1$ and consider nested bounded domains in $\R^3$ with smooth boundaries:
\[
 K_0\Subset K_1\Subset\cdots\Subset K_{n-1}\subset\R^3,
 \qquad \Sigma_i=\partial K_i,
 \qquad S=\bigcup_{i=0}^{n-1}\Sigma_i.
\]
Set
\[
 \Omega_0=K_0,\qquad
 \Omega_j=K_j\setminus\overline{K_{j-1}}\ (1\leq j<n),\qquad
 \Omega_n=\R^3\setminus\overline{K_{n-1}}.
\]
Given constants $0\leq v_0<v_1<\cdots<v_n$, let $\phi$ be the unique
continuous function on $\R^3$ satisfying
\begin{equation}\label{eq:profile}
 \Delta\phi=0\text{ on each }\Omega_j,\qquad
 \phi|_{\Sigma_i}=v_i,\qquad
 \lim_{|x|\to\infty}\phi(x)=v_n.
\end{equation}
Here $\Delta=\sum\limits_{k=1}^{3}\partial_k^2$ is the Laplacian.
Write $\phi_j=\phi|_{\Omega_j}$ and $f_j=*d\phi_j$; in particular,
$\phi_0=v_0$ and $f_0=0$.

Let $\nu_i$ be the unit outer normal to $\Sigma_i$.
Both one-sided normal derivatives below are taken in this same direction.
Define
\begin{equation}\label{eq:density}
 \sigma_i=\partial_{\nu_i}^{+}\phi-\partial_{\nu_i}^{-}\phi,
 \qquad
 d\mu=\sum\limits_{i=0}^{n-1}\sigma_i\,dS_i,
\end{equation}
where the signs $+$ and $-$ indicate the outer and inner traces.

\begin{definition}\label{def:bag}
The collection $\{(\Omega_j,f_j,\phi_j)\}_{j=0}^n$ is an
\emph{iterated magnetic bag}. We call it a
\emph{nonnegative-charge iterated magnetic bag} if $\sigma_i\geq0$
on every $\Sigma_i$. This is an additional condition on the domains and
the prescribed Higgs values $v_0,\ldots,v_n$.
\end{definition}

For a nonnegative-charge
iterated magnetic bag, the total magnetic charge is defined to be
\begin{equation}\label{eq:charge}
 q=\mu(S)=\sum\limits_{i=0}^{n-1}\int_{\Sigma_i}\sigma_i\,dS_i
   =\int_{\partial B_R}f_n>0
\end{equation}
for every sufficiently large $R$, with the sphere oriented outwards.
The function $\phi$ has the Newton-potential representation
\begin{equation}\label{eq:newton}
 \phi(x)=v_n-\frac1{4\pi}\int_S\frac{d\mu(y)}{|x-y|}.
\end{equation}

We equip $\su$ with the inner product
$\langle X,Y\rangle=-2\operatorname{tr}(XY)$.
This normalization may not be the most standard one, but it follows Jaffe and Taubes~\cite[Ch.~I, \S1]{JaffeTaubes1980}.
A smooth solution of the Bogomolny equation has mass $v>0$ and finite
energy if
\begin{equation}\label{eq:bogomolny}
 \lim_{|x|\to\infty}|\Phi(x)|=v,\qquad
 \int_{\R^3}\bigl(|F_{A,e}|^2+|d_{A,e}\Phi|^2\bigr)\,d^3x<\infty.
\end{equation}
By the results of Jaffe and Taubes~\cite[Ch.~IV]{JaffeTaubes1980}, the following
limit exists for every such monopole and determines its topological charge $N\in\mathbb Z$:
\begin{equation}\label{eq:monopole-charge}
 \lim_{R\to\infty}\int_{\partial B_R}
 \left\langle F_{A,e},\frac{\Phi}{|\Phi|}\right\rangle
 =\frac{4\pi N}{e}.
\end{equation}

\begin{theorem}[Main theorem]\label{thm:main}
Let $\{(\Omega_j,f_j,\phi_j)\}_{j=0}^n$ be a nonnegative-charge iterated magnetic bag
as in Definition~\ref{def:bag}, with continuous Higgs function $\phi$,
total charge $q>0$, and asymptotic Higgs value $v_n$.
There exist smooth finite-energy $SU(2)$ Bogomolny monopoles
$(A^{(k)},\Phi^{(k)})$ on $\R^3$, with couplings $e_k>0$, mass $v_n$,
and topological charges $N_k\to\infty$, such that
\begin{equation}\label{eq:scaling}
 e_k\longrightarrow\infty,\qquad
 \frac{4\pi N_k}{e_k}\longrightarrow q.
\end{equation}
Let $\F^{(k)}=*d|\Phi^{(k)}|$, viewed as a distributional two-form.
Away from the zeros of \(\Phi^{(k)}\), the Bogomolny equation implies that
\(\F^{(k)}\) coincides with the projected curvature
\(\langle F_{A^{(k)},e_k},\Phi^{(k)}/|\Phi^{(k)}|\rangle\).
Then the following hold:
\begin{enumerate}[label=\textup{(\roman*)},leftmargin=2em,itemsep=0.7em]
\item The Higgs norms satisfy $0\leq|\Phi^{(k)}|\leq v_n$ on $\R^3$ and
\[
 |\Phi^{(k)}|\longrightarrow\phi
 \quad\text{locally uniformly on }\R^3\setminus S.
\]
\item Let $U_+=\{x\in\R^3\setminus S:\phi(x)>0\}$.
For every $j=0,\ldots,n$ and every compact set $K\Subset\Omega_j\cap U_+$,
\[
 \F^{(k)}\longrightarrow f_j
 \quad\text{in }C^\infty(K).
\]
\item On all of $\R^3$,
\[
 |\Phi^{(k)}|\longrightarrow\phi\quad\text{in }L^1_{\mathrm{loc}},
 \qquad
 \F^{(k)}\rightharpoonup*d\phi
 \quad\text{as distributional two-forms}.
\]
\end{enumerate}
\end{theorem}

The nonnegative-charge condition is a limitation of the present paper,
but is also necessary in our setting: Any \(L^1_{\mathrm{loc}}\) limit
of the Higgs norms is subharmonic, and hence \(\Delta\phi=\mu\geq0\).
For Bolognesi's conjecture without the nonnegative-charge condition
(formulated appropriately), one possible direction is to study
the second-order Yang--Mills--Higgs equations by gluing monopoles and
antimonopoles. We hope to investigate this direction in future work.

The paper is organized as follows. Section~\ref{sec:gluing} constructs the
multi-center approximate solutions. Section~\ref{sec:arrays} discretizes
strictly positive surface charge densities into point arrays for the
multi-center approximation. Section~\ref{sec:deformation} deforms the
approximate solution into a genuine one. Section~\ref{sec:positive-limit}
completes the proof of the main theorem using a diagonal argument.

\section*{Acknowledgments}

Weifeng Sun thanks Siqi He for helpful discussions.
Weifeng Sun is supported by a start-up fund from Tsinghua University.
Wenhan Wang thanks Masashi Hamanaka and Gonçalo Oliveira for their interest in this work.

\section*{Disclosure of AI usage}

This work was completed with AI assistance. To be more precise, we (the authors)
originally focused on adapting Taubes' iterative scheme. Although we had also
vaguely considered a diagonal argument, we had not expected it to yield such
a simple proof. It was GPT-6 Astra that pointed out that the diagonal argument
could work without an iterative scheme.

All other key mathematical ideas and
arguments in the paper originated with the human authors, apart from
some auxiliary details generated by AI and carefully verified by the authors.

Besides GPT-6 Astra, the first author also used Gemini Argon and DeepSeek v4.1 1017
in polishing the
exposition and language and generating \LaTeX{} code under the authors' supervision.
In addition, in November 2026, the second author used Claude Fable 5.2
to explore results related to the paper, though none of these were directly
used in the main conclusions of this work.
The human authors take full responsibility for
all mathematical claims and for the writing of the paper.

\section{The approximate solution from gluing}\label{sec:gluing}

Our strategy to prove Theorem~\ref{thm:main} is first to construct a sequence
of approximate solutions, and then to correct them.
Define the Bogomolny residual of \(\Psi=(A,\Phi)\) by
\[
    g_\Psi:=*F_{A,e}-d_{A,e}\Phi .
\]

Let
\[
    E:=(\Lambda^1T^*\R^3\oplus\Lambda^0T^*\R^3)
       \otimes\operatorname{ad}P.
\]
On one-forms we use \(d_{A,e}^*=-*d_{A,e}*\). The gauge-fixed
linearized Bogomolny operator at \(\Psi\) is
\[
    D_\Psi(a,\varphi)
    :=
    \left(
        *d_{A,e}a-d_{A,e}\varphi+e[\Phi,a],
        \,
        d_{A,e}^*a-e[\Phi,\varphi]
    \right).
\]
Its formal \(L^2\)-adjoint is
\begin{equation}\label{eq:linearized-adjoint}
    D_\Psi^\dagger(b,\tau)
    =
    \left(
        *d_{A,e}b+d_{A,e}\tau-e[\Phi,b],
        \,
        -d_{A,e}^*b+e[\Phi,\tau]
    \right).
\end{equation}

We have the following Weitzenb\"ock identities in our conventions; compare Taubes~\cite{Taubes2013}.
For a smooth configuration \(\Psi=(A,\Phi)\) and \(u=(b,\tau)\in C_c^\infty(E)\),
\[
    D_\Psi D_\Psi^\dagger u
    =
    \nabla_A^*\nabla_Au
    +
    e^2[\Phi,[u,\Phi]]
    +
    V_\Psi u,
\]
where the zeroth-order endomorphism \(V_\Psi\) is given by
\[
    V_\Psi(b,\tau)
    :=
    e\left(
        *\bigl(g_\Psi\wedge b+b\wedge g_\Psi\bigr)+[g_\Psi,\tau],
        \,
        -*\bigl(*g_\Psi\wedge b-b\wedge *g_\Psi\bigr)
    \right).
\]
Here products of Lie-algebra-valued forms combine matrix and exterior multiplication.

A direct computation shows that:
\begin{itemize}
\item The endomorphism \(V_\Psi\) is self-adjoint, and
\begin{equation}\label{eq:weitzenbock-energy}
    \|D_\Psi^\dagger u\|_{L^2}^2
    =
    \|\nabla_Au\|_{L^2}^2
    +
    e^2\|[\Phi,u]\|_{L^2}^2
    +
    \int_{\R^3}
    \langle u,V_\Psi u\rangle\,d^3x .
\end{equation}

\item There is a constant \(C>0\), independent of \(e\), \(\Psi\), and \(x\in\R^3\), such that
\begin{equation}\label{eq:weitzenbock-residual-bound}
    |V_\Psi(x)|_{\mathrm{op}}\leq C e|g_\Psi(x)|.
\end{equation}
\end{itemize}

Let \(\Psi=(A,\Phi)\) be a smooth configuration satisfying \eqref{eq:bogomolny}
for some \(v>0\), with \(|\Phi|\leq v\) on \(\R^3\).
For sufficiently large \(R\), define the topological charge \(N\) of \(\Psi\) by
\[
 N:=\deg\!\left(\frac{\Phi}{|\Phi|}\Big|_{S_R^2}\right).
\]
This definition is independent of the choice of such \(R\).
If \(\Psi\) solves the Bogomolny equation, this degree agrees with the
topological charge \(N\) in formula~\eqref{eq:monopole-charge}.
For \(u\in C_c^\infty(E)\), set
\[
 \|u\|_{H_{\Psi,e}}^2
 :=\|\nabla_Au\|_{L^2}^2+e^2\|[\Phi,u]\|_{L^2}^2,
\]
and let \(H_{\Psi,e}\) be the Hilbert completion in this norm.
Define the relative residual functional by
\[
 R(\Psi):=\sup_{0\neq u\in C_c^\infty(E)}
 \frac{\left|\int_{\R^3}\langle u,V_\Psi u\rangle\,d^3x\right|}
 {\|u\|_{H_{\Psi,e}}^2}.
\]
We call \(\Psi=(A,\Phi)\) an \emph{admissible approximate solution} if, in
addition, \(R(\Psi)<1\). Then
\[
 (1-R(\Psi))\|u\|_{H_{\Psi,e}}^2
 \leq\|D_\Psi^\dagger u\|_{L^2}^2
 \leq(1+R(\Psi))\|u\|_{H_{\Psi,e}}^2
 \qquad(u\in C_c^\infty(E)).
\]
In particular, \(D_\Psi^\dagger\) extends to a bounded map
\(H_{\Psi,e}\to L^2(E)\).

\begin{proposition}\label{prop:linear-right-inverse-coercive}
Let \(\Psi=(A,\Phi)\) be an admissible approximate solution.
For every \(w\in L^{6/5}(E)\), there is a unique
\(u_w\in H_{\Psi,e}\) with \(D_\Psi D_\Psi^\dagger u_w=w\) in the weak sense, and
\[
 P_\Psi w:=D_\Psi^\dagger u_w\in L^2(E),\qquad D_\Psi P_\Psi w=w
\]
distributionally. Moreover,
\[
 \|u_w\|_{H_{\Psi,e}}\leq\frac{\|w\|_{L^{6/5}}}{1-R(\Psi)},\qquad
 \|P_\Psi w\|_{L^2}\leq\frac{\|w\|_{L^{6/5}}}{\sqrt{1-R(\Psi)}}.
\]
\end{proposition}

\begin{proof}
Define the bilinear form
\[
 B(u,z):=\int_{\R^3}\langle D_\Psi^\dagger u,D_\Psi^\dagger z\rangle\,d^3x.
\]
Then \(B\) is bounded, and since \(R(\Psi)<1\), the lower bound shows that
\(B\) is also coercive.
By the Sobolev and Kato inequalities, we obtain \(\|z\|_{L^6}\leq\|z\|_{H_{\Psi,e}}\).
Thus \(z\mapsto\displaystyle\int_{\R^3}\langle w,z\rangle\,d^3x\) is bounded on \(H_{\Psi,e}\)
with norm at most \(\|w\|_{L^{6/5}}\). All conclusions follow directly
from the Lax--Milgram theorem.
\end{proof}

We recall the explicit formula for the charge-one Prasad--Sommerfield
monopole~\cite{PrasadSommerfield1975}.
Throughout the rest of this section, the asymptotic Higgs norm \(v>0\) is
fixed.

Fix an orthonormal basis \(\{\tau_1,\tau_2,\tau_3\}\) of \(\su\), with
\begin{equation}\label{eq:core-0}
 [\tau_i,\tau_j]=-\varepsilon_{ijk}\tau_k.
\end{equation}
Let
\[
    r:=|x|,
    \qquad
    n_i:=\frac{x_i}{r}.
\]
For \(e,m>0\), the Prasad--Sommerfield monopole centered at the origin is
\[
    \Psi^{\mathrm{PS}}_{e,m}
    =
    \bigl(A^{\mathrm{PS}}_{e,m},\Phi^{\mathrm{PS}}_{e,m}\bigr),
\]
where, on \(\R^3\setminus\{0\}\),
\[
    A^{\mathrm{PS}}_{e,m}
    =
    \left(
        \frac{1}{er}
        -
        \frac{m}{\sinh(emr)}
    \right)
    \varepsilon_{ijk}n_i\,dx^j\,\tau_k,
    \qquad
    \Phi^{\mathrm{PS}}_{e,m}
    =
    \left(
        m\coth(emr)
        -
        \frac{1}{er}
    \right)n_i\tau_i.
\]
The apparent singularity at the origin is removable. The resulting smooth
configuration satisfies
\[
    F_{A^{\mathrm{PS}}_{e,m},e}
    =
    *d_{A^{\mathrm{PS}}_{e,m},e}\Phi^{\mathrm{PS}}_{e,m}
\]
and has asymptotic Higgs norm \(m\) and topological charge one.
Its Higgs field \(\Phi^{\mathrm{PS}}_{e,m}\) vanishes only at the origin.
We denote the translate of this monopole to the point \(p\) by \(\Psi^{\mathrm{PS}}_{e,m,p}\).

We use the following reducible multi-center Dirac background.

\begin{definition}
Let
\[
    \Theta:=\{p_1,\ldots,p_N\}\subset\R^3
\]
be a set of distinct points. Define
\[
    \phi_\Theta(x)
    =
    v-\frac1e\sum\limits_{j=1}^{N}\frac1{|x-p_j|}.
\]
The reducible charge-\(N\) Dirac background on
\(\R^3\setminus\Theta\) is specified, up to reducible gauge, by a unit
adjoint section \(\sigma_\Theta\) and a connection \(A_\Theta\)
on a principal \(SU(2)\)-bundle, satisfying
\[
    d_{A_\Theta,e}\sigma_\Theta=0,
    \qquad
    F_{A_\Theta,e}
    =
    (*d\phi_\Theta)\sigma_\Theta.
\]
Indeed, on every positively oriented sphere \(S_r^2(p_i)\) enclosing only
\(p_i\),
\[
    \int_{S_r^2(p_i)}*d\phi_\Theta
    =
    \frac{4\pi}{e},
\]
which corresponds to topological charge one in our normalization. Set
\[
    \Phi_\Theta
    :=
    \phi_\Theta\sigma_\Theta,
    \qquad
    \Psi_\Theta
    :=
    (A_\Theta,\Phi_\Theta).
\]
Then, explicitly,
\[
    F_{A_\Theta,e}
    =
    *d_{A_\Theta,e}\Phi_\Theta
    \qquad\text{on }\R^3\setminus\Theta.
\]
\end{definition}

For \(p\in\R^3\), write
\[
    r_p:=|x-p|,
    \qquad
    n_{p,i}:=\frac{x_i-p_i}{r_p},
    \qquad
    \sigma_p^{\mathrm D}:=n_{p,i}\tau_i.
\]
For \(\Theta=\{p\}\) and \(v=m>0\), we call it a charge-one Dirac monopole,
presented in the hedgehog gauge as follows:
\[
 A^{\mathrm D}_{e,m,p}
 =\frac1{er_p}\varepsilon_{ijk}n_{p,i}\,dx^j\,\tau_k,
 \qquad
 \Phi^{\mathrm D}_{e,m,p}
 =\left(m-\frac1{er_p}\right)\sigma_p^{\mathrm D}
 \qquad\text{on }\R^3\setminus\{p\}.
\]

Define the effective mass at the \(i\)-th center by
\[
    m_i
    :=
    v-\frac1e\sum\limits_{j\neq i}\frac1{|p_i-p_j|}.
\]
For \(e,m,L>0\) and \(p\in\R^3\), define the \(L\)-shell of the monopole
\[
    \mathcal S_L^{e,m}(p)
    :=
    \left\{
        x\in\R^3\colon
        \frac{3L}{em}
        \leq
        |x-p|
        \leq
        \frac{6L}{em}
    \right\}.
\]
For each \(i\), set
\[
    \mathcal S_i
    :=
    \mathcal S_L^{e,m_i}(p_i).
\]
We call the inner balls \(B_{3L/(em_i)}(p_i)\) the cores.

\begin{definition}
\label{def:admissible-lower-mass-condition}
Let \(e,v>0\), \(L>1\), and
\(\Theta=\{p_1,\ldots,p_N\}\). Define
\[
    S_2(\Theta)
    :=
    \max_{1\leq i\leq N}
    \sum\limits_{j\neq i}\frac1{|p_i-p_j|^2}.
\]
We say that \((\Theta,L)\) satisfies the \emph{admissible mass--separation
condition} with dimensionless lower mass ratio \(r_*\in(0,1)\) if
\[
    m_i\geq r_*v,\qquad i=1,\ldots,N,
    \qquad
    C_{r_*}\frac{L S_2(\Theta)}{e^2v^2}\leq\frac12
    \left(1-\frac1{3L}\right),
\]
and
\[
    |p_i-p_j|
    \geq
    \left(1+\frac{12}{r_*}\right)
    \frac{L}{ev},
    \qquad i\neq j.
\]
Here \(C_{r_*}\geq24/r_*^2\) is a fixed constant depending
only on \(r_*\).

\end{definition}

For each \(i\), write \(r_i=|x-p_i|\) and split the Dirac Higgs function as
\begin{equation}\label{eq:local-multidirac-Higgs-splitting}
 \phi_\Theta(x)=m_i-\frac1{er_i}+\eta_i(x),\qquad
 \eta_i(x)=-\frac1e\sum_{j\ne i}
 \left(\frac1{|x-p_j|}-\frac1{|p_i-p_j|}\right).
\end{equation}
Thus \(\eta_i(p_i)=0\), and \(\eta_i\) is harmonic away from the other centers.

\begin{lemma}\label{lem:lower-mass-nonvanishing-on-shells}
Assume that \((\Theta,L)\) satisfies the admissible mass--separation
condition. Then the shells \(\mathcal S_i\) are pairwise disjoint and, for
\(x\in\mathcal S_i\),
\[
    \phi_\Theta(x)
    \geq
    \frac{m_i}{2}
    \left(1-\frac1{3L}\right)
    >0.
\]
In particular, \(\Phi_\Theta/|\Phi_\Theta|\) is defined on every shell.

\end{lemma}

\begin{proof}
The mass and separation bounds give
\[
 |p_i-p_j|-\frac{6L}{em_i}-\frac{6L}{em_j}\geq\frac{L}{ev}>0
 \qquad(i\ne j),
\]
so the shells are disjoint. On \(B_{6L/(em_i)}(p_i)\), separation also gives
\(|x-p_j|\geq |p_i-p_j|/2\) for \(j\ne i\). Differentiating the Newton
kernel \(|x-y|^{-1}\) gives \(|d\eta_i|\leq4S_2(\Theta)/e\).
Integrating from \(p_i\) to \(x\) yields
\[
 |\eta_i(x)|\leq\frac{4r_i}{e}S_2(\Theta)
 \leq\frac{24L}{e^2m_i}S_2(\Theta)
 \leq\frac{m_i}{2}\left(1-\frac1{3L}\right).
\]
The last inequality follows from admissibility, with
\(C_{r_*}\geq24/r_*^2\). Since \(|x-p_i|\geq3L/(em_i)\),
\[
 \phi_\Theta(x)\geq m_i-\frac{m_i}{3L}
 -\frac{m_i}{2}\left(1-\frac1{3L}\right)
 =\frac{m_i}{2}\left(1-\frac1{3L}\right)>0.
\]
\end{proof}

We follow Taubes' construction of approximate monopoles by gluing
Prasad--Sommerfield monopoles to a Dirac background~\cite{Taubes2013}.
Write \(\sigma_i^{\mathrm D}=\sigma_{p_i}^{\mathrm D}\).

Since \(\eta_i\) is harmonic, \(*d\eta_i\) is closed. By the radial
Poincar\'e lemma, choose a smooth one-form \(\omega_i\)
on \(B_{6L/(em_i)}(p_i)\) such that
\[
 d\omega_i=*d\eta_i,\qquad
 \iota_{\partial_{r_i}}\omega_i=0,\qquad
 \omega_i(p_i)=0.
\]
For the charge-one Dirac monopole centered at \(p_i\), write \(A_i^{\mathrm D}=A_{e,m_i,p_i}^{\mathrm D}\)
and \(\Phi_i^{\mathrm D}=\Phi_{e,m_i,p_i}^{\mathrm D}\).
Because the Dirac connection preserves \(\sigma_i^{\mathrm D}\),
\[
 F_{A_i^{\mathrm D}+\omega_i\sigma_i^{\mathrm D},e}
 =(*d\phi_\Theta)\sigma_i^{\mathrm D},
 \qquad
 d_{A_i^{\mathrm D}+\omega_i\sigma_i^{\mathrm D},e}
       (\phi_\Theta\sigma_i^{\mathrm D})
 =(d\phi_\Theta)\sigma_i^{\mathrm D}.
\]
Thus \((A_i^{\mathrm D}+\omega_i\sigma_i^{\mathrm D},
\phi_\Theta\sigma_i^{\mathrm D})\) is the local reducible Bogomolny
model for the multi-center Dirac background near \(p_i\).

Recall that on \(B_{6L/(em_i)}(p_i)\setminus\{p_i\}\), the Dirac background
and the local model
\((A_i^{\mathrm D}+\omega_i\sigma_i^{\mathrm D},\phi_\Theta\sigma_i^{\mathrm D})\)
have the same local charge, and their induced \(U(1)\) connections have the same curvature.
After identifying the corresponding charge-one \(U(1)\) bundles, the difference of their abelian
connections is closed and hence exact on this simply connected punctured
ball. We therefore fix a common local gauge in which
\begin{equation}
 \sigma_\Theta=\sigma_i^{\mathrm D},
 \qquad
 A_\Theta=A_i^{\mathrm D}+\omega_i\sigma_i^{\mathrm D},
 \qquad
 \Phi_\Theta=\Phi_i^{\mathrm D}+\eta_i\sigma_i^{\mathrm D}.
\label{eq:fixed-local-hedgehog-identification}
\end{equation}
Write \(\Psi_i^{\mathrm{PS}}=(A_i^{\mathrm{PS}},\Phi_i^{\mathrm{PS}})\)
for \(\Psi_{e,m_i,p_i}^{\mathrm{PS}}\) in this gauge.
On \(\mathcal S_i\), Lemma~\ref{lem:lower-mass-nonvanishing-on-shells}
and the explicit Prasad--Sommerfield formula give
\[
 \frac{\Phi_\Theta}{|\Phi_\Theta|}
 =\sigma_i^{\mathrm D}
 =\frac{\Phi_i^{\mathrm{PS}}}{|\Phi_i^{\mathrm{PS}}|}.
\]
All differences and cutoff interpolations below are taken in these
fixed local gauges.
Here longitudinal and transverse refer to the components parallel and
orthogonal to \(\sigma_i^{\mathrm D}\), respectively.

\begin{lemma}
\label{lem:local-ps-dirac-comparison-on-shell}
Assume the admissible mass--separation condition. For every integer \(\ell\geq0\),
there are constants \(C_{r_*,\ell}>0\) and \(c_{\ell}>0\), independent of
\(e,N,L,\Theta\), and \(i\), such that on \(\mathcal S_i\),
\[
 \left| \nabla_{A_\Theta}^{\ell} (A_\Theta-A_i^{\mathrm{PS}})\right|
 +\left| \nabla_{A_\Theta}^{\ell} (\Phi_\Theta-\Phi_i^{\mathrm{PS}})\right|
 \leq C_{r_*,\ell}\left(\frac{em_i}{L}\right)^{\ell}
 \left( \frac{L S_2(\Theta)}{e^2m_i} + m_i \exp(-c_{\ell}L) \right).
\]
On \(\mathcal S_i\), the transverse components satisfy
\[
 \left|(A_\Theta-A_i^{\mathrm{PS}})_\perp\right|
 \leq C_{r_*} m_i \exp(-c_0L),
 \qquad
 (\Phi_\Theta-\Phi_i^{\mathrm{PS}})_\perp=0.
\]

\end{lemma}

\begin{proof}
Work in the local gauge fixed in \eqref{eq:fixed-local-hedgehog-identification}.
On \(B_{6L/(em_i)}(p_i)\), the preceding Newton-kernel estimate bounds
\(|d\eta_i|\) by \(4S_2(\Theta)/e\) and \(|\eta_i|\) by this bound times
the ball radius. By the separation assumption on the centers,
\(|x-p_j|\geq 6L/(em_i)\) on this ball for \(j\ne i\).
Thus each additional derivative of the Newton kernel contributes a factor
of order \(em_i/L\) to the estimate.
The radial Poincar\'e estimate likewise bounds \(\omega_i\)
by this radius times the supremum of \(|d\eta_i|\). For \(\ell\geq1\),
its derivative estimates take the form
\[
 \|\partial^\ell\omega_i\|_{L^\infty}
 \leq C_\ell\left(
 \frac{L}{em_i}\|\partial^{\ell+1}\eta_i\|_{L^\infty}
 +\|\partial^\ell\eta_i\|_{L^\infty}\right),
\]
where all norms are taken over \(B_{6L/(em_i)}(p_i)\).
The preceding Newton-kernel bounds therefore give the same scaling
for both terms, yielding, for every \(\ell\geq0\),
\[
 |\partial^{\ell}\eta_i|+|\partial^{\ell}\omega_i|
 \leq C_{r_*,\ell}\left(\frac{em_i}{L}\right)^{\ell}
       \frac{LS_2(\Theta)}{e^2m_i}.
\]
Since both the Dirac and background connections preserve
\(\sigma_i^{\mathrm D}\), the same bound holds for the covariant
derivatives of \(\eta_i\sigma_i^{\mathrm D}\) and
\(\omega_i\sigma_i^{\mathrm D}\).

Using the decompositions
\[
 A_\Theta-A_i^{\mathrm{PS}}
 =\omega_i\sigma_i^{\mathrm D}+(A_i^{\mathrm D}-A_i^{\mathrm{PS}}),
 \qquad
 \Phi_\Theta-\Phi_i^{\mathrm{PS}}
 =\eta_i\sigma_i^{\mathrm D}
   +(\Phi_i^{\mathrm D}-\Phi_i^{\mathrm{PS}}),
\]
it remains to estimate the differences between the Dirac and
Prasad--Sommerfield fields at \(p_i\). The explicit formulas in the
same hedgehog gauge give
\[
 A_i^{\mathrm{PS}}-A_i^{\mathrm D}
 =-\frac{m_i}{\sinh(em_ir_i)}
   \varepsilon_{abc}n_{p_i,a}\,dx^b\,\tau_c,
 \qquad
 \Phi_i^{\mathrm{PS}}-\Phi_i^{\mathrm D}
 =m_i\bigl(\coth(em_ir_i)-1\bigr)\sigma_i^{\mathrm D}.
\]
Write \(\nabla^{\mathrm D}=\nabla_{A_i^{\mathrm D}}\).
On \(\mathcal S_i\), direct calculation using the above explicit formula leads to
\[
 |(\nabla^{\mathrm D})^\ell(A_i^{\mathrm{PS}}-A_i^{\mathrm D})|
 +|(\nabla^{\mathrm D})^\ell(\Phi_i^{\mathrm{PS}}-\Phi_i^{\mathrm D})|
 \leq C_\ell\left(\frac{em_i}{L}\right)^\ell m_i \exp(-c_\ell L),
\]
where \(C_\ell,c_\ell>0\) depend only on \(\ell\).
Together with \eqref{eq:fixed-local-hedgehog-identification} and the
preceding remainder bounds, this gives the first estimate, with the
polynomial factors in \(L\) absorbed into the exponential.

The remainder terms and the single-center Higgs difference are
longitudinal. Thus the transverse Higgs difference is zero, while the
transverse connection difference bound is given by the case \(\ell=0\)
of the single-center estimate above.
These give the second estimate.
\end{proof}

Choose a smooth cutoff \(\chi:[0,\infty)\to[0,1]\) equal to \(1\) on
\([0,3]\) and to \(0\) on \([6,\infty)\). Set
\[
    \chi_i(x):=\chi\!\left(\frac{em_i|x-p_i|}{L}\right),
    \qquad \operatorname{supp}(d\chi_i)\subset\mathcal S_i.
\]

Glue the Prasad--Sommerfield principal bundles over the balls
\(B_{6L/(em_i)}(p_i)\) to the exterior Dirac bundle over
\[
    \R^3\setminus
    \bigcup_{i=1}^{N}\overline{B_{3L/(em_i)}(p_i)}
\]
using the local gauge identifications fixed above on
\(\mathcal S_i\). In these local gauges, define the glued configuration
\(\Psi_{\mathrm{app}}=(A_{\mathrm{app}},\Phi_{\mathrm{app}})\) by
\[
    \Psi_{\mathrm{app}}(x)
    :=
    \begin{cases}
        \Psi_i^{\mathrm{PS}}(x),
        & |x-p_i|\leq\frac{3L}{em_i},\\
        \chi_i(x)\Psi_i^{\mathrm{PS}}(x)
        +(1-\chi_i(x))\Psi_\Theta(x),
        & x\in\mathcal S_i,\\
        \Psi_\Theta(x),
        & x\notin\displaystyle\bigcup_{i=1}^N B_{6L/(em_i)}(p_i).
    \end{cases}
\]
Its residual is
\[
    g_{\mathrm{app}}
    :=
    *F_{A_{\mathrm{app}},e}
    -
    d_{A_{\mathrm{app}},e}
    \Phi_{\mathrm{app}}.
\]
Assume the admissible mass--separation condition. Then
\(\Psi_{\mathrm{app}}\) is a globally smooth configuration on
the glued bundle satisfying \eqref{eq:bogomolny}, with topological charge \(N\).
Moreover,
\[
    0\leq|\Phi_{\mathrm{app}}|\leq v
    \qquad\text{on }\R^3,
\]
and
\[
    \{\Phi_{\mathrm{app}}=0\}
    =
    \{p_1,\ldots,p_N\}.
\]

\begin{proposition}
\label{prop:glued-approximate-basic-estimates}
Under the admissible mass--separation condition, on every shell
\(\mathcal S_i\), \(\Phi_{\mathrm{app}}\) is parallel to
\(\Phi_\Theta/|\Phi_\Theta|\), and
\[
    |\Phi_{\mathrm{app}}|
    \geq
    \frac{m_i}{2}
    \left(1-\frac1{3L}\right)
    \geq
    \frac{r_* v}{3}.
\]
Moreover,
\[
    \operatorname{supp}g_{\mathrm{app}}
    \subset
    \bigcup_{i=1}^{N}\mathcal S_i,
\]
so the residual vanishes on an open neighborhood of the Higgs zero set.
On every shell \(\mathcal S_i\),
\[
 |g_{\mathrm{app}}|
 \leq C_{r_*,\chi}\left(
       \frac{S_2(\Theta)}e+ev^2\exp(-cL)\right),
\]
where \(C_{r_*,\chi},c>0\) are independent of \(e,N,L,\Theta\), and \(i\).

\end{proposition}

\begin{proof}
On \(\mathcal S_i\), the Prasad--Sommerfield formula gives
\[
 |\Phi_i^{\mathrm{PS}}|
 \geq m_i\left(1-\frac1{3L}\right).
\]
The Dirac Higgs function \(\phi_\Theta\) is positive on these shells by
Lemma~\ref{lem:lower-mass-nonvanishing-on-shells}, and hence positive
throughout the exterior by the minimum principle and its limit \(v\)
at infinity; its explicit formula also gives \(\phi_\Theta<v\).
The aligned Higgs directions on each shell yield
\[
 |\Phi_{\mathrm{app}}|
 =\chi_i|\Phi_i^{\mathrm{PS}}|+(1-\chi_i)\phi_\Theta
 \geq\frac{m_i}{2}\left(1-\frac1{3L}\right)
 \geq\frac{r_* v}{3}.
\]
The configuration solves the Bogomolny
equation off the closed shells, so its residual is supported in those
shells and vanishes on the open cores containing all Higgs zeros.

On \(\mathcal S_i\), put
\[
 \alpha_i=A_\Theta-A_i^{\mathrm{PS}},\qquad
 \beta_i=\Phi_\Theta-\Phi_i^{\mathrm{PS}},\qquad t_i=1-\chi_i.
\]
Since both \(\Psi_i^{\mathrm{PS}}\) and \(\Psi_\Theta\) solve the Bogomolny
equation, we obtain the exact identity
\begin{equation}\label{eq:exact-cutoff-residual}
 g_{\mathrm{app}}
 =*(dt_i\wedge\alpha_i)-dt_i\,\beta_i
 +et_i(t_i-1)\bigl(*(\alpha_i\wedge\alpha_i)-[\alpha_i,\beta_i]\bigr).
\end{equation}
Since \(|dt_i|\leq C_\chi em_i/L\),
Lemma~\ref{lem:local-ps-dirac-comparison-on-shell} bounds the linear
terms by
\[
 C_{r_*,\chi}\left(\frac{S_2(\Theta)}e+em_i^2\exp(-cL)\right).
\]
For the quadratic terms, longitudinal factors commute, so each nonzero
bracket contains a transverse connection factor, bounded by
\(C_{r_*}m_i \exp(-cL)\) by the same lemma. Admissibility gives
\(LS_2(\Theta)/(e^2m_i^2)\leq C_{r_*}\), hence
\(|\alpha_i|+|\beta_i|\leq C_{r_*}m_i\). The quadratic terms are therefore
bounded by \(C_{r_*}em_i^2\exp(-cL)\). Combining these estimates and
using \(m_i\leq v\) proves the residual bound.
\end{proof}

Thus the glued configuration \(\Psi_{\mathrm{app}}\) is an approximate
solution of the Bogomolny equation, with residual controlled by
Proposition~\ref{prop:glued-approximate-basic-estimates}.
Before deforming it into an exact solution, we first construct in the
next section point arrays that approximate the prescribed surface charge
and satisfy the mass and separation conditions needed for gluing.

\section{Discretization of the surface charge}\label{sec:arrays}

The purpose of this section is
to construct a sequence of point arrays \(\Theta\) whose associated Dirac Higgs
functions converge to the Higgs function of the prescribed magnetic bag.
Given such point arrays, the multi-center approximate monopoles constructed in the previous section
will then approximate the prescribed magnetic bag.
These points should also be well separated to allow the approximate
monopoles to be deformed into genuine ones later.

Recall that the prescribed nonnegative-charge iterated magnetic bag
\(\{(\Omega_j,f_j,\phi_j)\}_{j=0}^n\) has nonnegative surface charge densities
\[
 \sigma_i\geq0.
\]
For a finite signed measure \(\lambda\) on \(S\), write
\[
 U_\lambda(x)=\int_S\frac{d\lambda(y)}{|x-y|}.
\]
Then the Newton-potential representation \eqref{eq:newton} becomes
\[
 \phi=v_n-\frac1{4\pi}U_\mu.
\]
At the level of measures, the aim is to choose points \(x_{a,N}\in S\)
such that
\[
 \frac qN\sum\limits_{a=1}^N\delta_{x_{a,N}}
 \rightharpoonup\mu
 \qquad\text{weakly as }N\to\infty,
\]
where \(\mu\) is the surface charge measure
\(d\mu=\sum\limits_{i=0}^{n-1}\sigma_i\,dS_i\), \(q=\mu(S)\) is its total charge,
and \(\delta_x\) is the Dirac measure at \(x\).
The points must also be well separated so that the gluing construction
of Section~\ref{sec:gluing} applies.
In this section, we restrict ourselves to the case where
\(\sigma_i>0\) on every \(\Sigma_i\).

\begin{theorem}[Main theorem of Section~\ref{sec:arrays}]\label{thm:arrays-normalized-arrays}
Let \(\nu\) be a probability measure on \(S\) with smooth, strictly
positive densities on all connected components. Let \(v,q_0>0\), and suppose
that
\[
 \psi(x)=v-\frac{q_0}{4\pi}U_\nu(x)\geq\rho v\qquad\text{on }\R^3
\]
for some \(\rho>0\). For every sufficiently large \(N\), there exist
distinct points \(\Theta_N=\{x_{1,N},\ldots,x_{N,N}\}\subset S\) with the
following properties. Rescale the point arrays by setting
\[
 \lambda_N=\frac{4\pi Nv}{q_0},\qquad
 p_{a,N}=\lambda_Nx_{a,N},\qquad
 \lambda_N\Theta_N=\{p_{1,N},\ldots,p_{N,N}\}.
\]
We also write \(\lambda_NS=\{\lambda_Nx:x\in S\}\).
Define the Dirac Higgs function and the effective masses by
\[
 \phi_N^{\mathrm D}(z)=1-\sum\limits_{a=1}^N|z-p_{a,N}|^{-1},\qquad
 m_{a,N}=1-\sum\limits_{a'\neq a}|p_{a,N}-p_{a',N}|^{-1}.
\]
Then the following hold.
\begin{itemize}
\item We have
\[
 p_{a,N}\in\lambda_NS\cap B(0,CN),\qquad
 \min_{a\neq b}|p_{a,N}-p_{b,N}|\geq c\sqrt N.
\]
\item The effective masses satisfy
\[
 \rho/2\leq m_{a,N}\leq1.
\]
\item
For every integer \(j\geq1\), if \(p_{a_*,N}\) is a point in
\(\lambda_N\Theta_N\) closest to \(z\), then
\[
 \sum\limits_{a\neq a_*}|z-p_{a,N}|^{-j}\leq
 \begin{cases}
 C,&j=1,\\
 C(\log N)/N,&j=2,\\
 C_jN^{-j/2},&j\geq3.
 \end{cases}
\]
In particular, taking \(z=p_{a,N}\) and \(j=2\), we obtain
\[
 S_2(\lambda_N\Theta_N)
 =\max_a\sum\limits_{a'\neq a}|p_{a,N}-p_{a',N}|^{-2}
 \leq C\frac{\log N}{N}.
\]
\item
For every $r\geq0$ and every compact $K\Subset\R^3\setminus S$,
\[
 v\phi_N^{\mathrm D}(\lambda_Nx)
 =v-\frac{q_0}{4\pi N}\sum\limits_a|x-x_{a,N}|^{-1}
 \longrightarrow\psi(x)
 \qquad\text{in }C^r(K).
\]
\end{itemize}
All constants are independent of \(N\).
\end{theorem}

Let \(S_1,\ldots,S_B\) be the connected components of \(S\).
We equip $S_b$ with its
intrinsic (geodesic) distance $d_b$ and write
\[
 B_{d_b}(x,r)=\{y\in S_b:d_b(x,y)<r\}
\]
for intrinsic balls, while $B(z,r)$ always denotes a Euclidean ball in
$\R^3$. Since $|x-y|\leq d_b(x,y)$, we have $B_{d_b}(x,r)\subset B(x,r)$.

\begin{lemma}\label{lem:arrays-geometric-potential-preliminaries}
Let $\lambda$ be a signed measure on $S$ of the form
\[
 d\lambda=\sum\limits_{b=1}^{B}\omega_b\,dS_b
\]
with bounded densities $\omega_b$. The following hold, with constants
depending only on $S$ and on the densities.
\begin{enumerate}[label=\textup{(\arabic*)}]
\item For all $z\in\R^3$ and $0<r\leq1$,
  \[
   |\lambda|\bigl(B(z,r)\bigr)\leq Cr^2
   \qquad\text{and}\qquad
   \int_{S\cap B(z,r)}\frac{d|\lambda|(y)}{|z-y|}\leq Cr.
  \]
\item The potential $U_\lambda$ is bounded and continuous on $\R^3$ and
  smooth on $\R^3\setminus S$, and derivatives of every order may be taken
  under the integral sign on compact subsets of $\R^3\setminus S$.
  Moreover,
  \[
   U_\lambda(x)=\int_S\frac{d\lambda(y)}{|x-y|}
   =\frac{\lambda(S)}{|x|}+O(|x|^{-2})
   \qquad\text{as }|x|\to\infty.
  \]
\item There is a constant $C$ depending only on $S$ such that
  \[
   |x-y|\leq d_b(x,y)\leq C|x-y|
   \qquad(x,y\in S_b,\ 1\leq b\leq B).
  \]
\item If $\omega_b$ is smooth and strictly positive on $S_b$, then
  $(S_b,d_b,\omega_b\,dS_b)$ is a compact connected $2$-Ahlfors regular
  metric measure space; that is, there are constants $0<c\leq C$ with
  \[
   cr^2\leq\int_{B_{d_b}(x,r)}\omega_b\,dS_b\leq Cr^2,
   \qquad x\in S_b,\ 0<r\leq\operatorname{diam}_{d_b}S_b.
  \]
  The same holds after multiplying the measure by any positive constant.
\end{enumerate}
\end{lemma}

\begin{proof}
All assertions except (2) follow directly from straightforward estimates
in smooth local coordinate charts and compactness.
For the second assertion of (1), summing over dyadic annuli gives
\[
 \int_{S\cap B(z,r)}\frac{d|\lambda|(y)}{|z-y|}
 \leq\sum\limits_{j\geq0}\frac{2^{j+1}}r\,C(2^{-j}r)^2\leq Cr.
\]
For (2), split the potential into a near part, uniformly \(O(r)\) by (1),
and a nonsingular far part to obtain boundedness and continuity.
The asymptotics follow from the uniform
expansion \(|x-y|^{-1}=|x|^{-1}+O(|x|^{-2})\) for \(y\in S\).
\end{proof}

We now construct the point arrays.

\begin{lemma}\label{lem:arrays-componentwise-equal-mass-cells}
Let
\[
 d\nu=\sum\limits_{b=1}^{B}\omega_b\,dS_b
\]
be a probability measure on $S$ whose densities $\omega_b$ are smooth and
strictly positive.
For every sufficiently large $N$ there are positive integers
$N_1,\dots,N_B$ with
\[
 \sum\limits_{b=1}^BN_b=N,\qquad |N_b-N\nu(S_b)|\leq1,
\]
and, for each $b$, a partition of $S_b$ into $N_b$ measurable sets together
with a representative point in each set, with the following properties.
Enumerate all sets in the partitions and their representatives by a single index
$a\in\{1,\dots,N\}$ as $P_a$ and $x_a=x_{a,N}$, suppressing the index
$N$ on the points in this and the following lemmas. Put
\[
 \Theta_N=\{x_1,\dots,x_N\}\subset S,\qquad h=N^{-1/2}.
\]
Then there is a constant $C$, independent of $N$, such that:
\begin{enumerate}[label=\textup{(\arabic*)}]
\item If $P_a\subset S_b$, then
  \[
   \nu(P_a)=\frac{\nu(S_b)}{N_b}=\frac1N+O(N^{-2}),\qquad
   \operatorname{diam}_{d_b}P_a\leq CN_b^{-1/2},\qquad
   B_{d_b}\bigl(x_a,C^{-1}N_b^{-1/2}\bigr)\subset P_a;
  \]
  in particular, there is a constant $C_1$, independent of $N$, with
  \begin{equation}\label{eq:cell-euclid-diam}
   P_a\subset B(x_a,C_1h)
   \qquad\text{and}\qquad
   B_{d_b}(x_a,C_1^{-1}h)\subset P_a;
  \end{equation}
\item $|x_a-x_{a'}|\geq C^{-1}h$ for all $a\neq a'$;
\item $\#\bigl(\Theta_N\cap B(z,R)\bigr)\leq C(1+NR^2)$ for all $z\in\R^3$ and
  $R>0$.
\end{enumerate}
\end{lemma}

\begin{proof}
Rounding \(N\nu(S_b)\) down and distributing the remaining units gives
the required integers \(N_b\). We write \(a_N\asymp b_N\) if
\(c b_N\leq a_N\leq C b_N\) for positive constants independent of \(N\).
Since \(\nu(S_b)>0\), we have
\(N_b\asymp N\) and \(\nu(S_b)/N_b=1/N+O(N^{-2})\).

By Lemma~\ref{lem:arrays-geometric-potential-preliminaries}(4), each
\((S_b,d_b,\nu|_{S_b}/\nu(S_b))\) is a compact connected \(2\)-Ahlfors
regular metric measure space. The equal-measure partition theorem of
Gigante and Leopardi~\cite{GiganteLeopardi} therefore gives a partition into \(N_b\) sets
of \(\nu\)-measure \(\nu(S_b)/N_b\), each of diameter at most
\(CN_b^{-1/2}\) and containing an intrinsic ball of radius
\(cN_b^{-1/2}\). Choosing the centers of these inner balls as representatives
proves (1).

On the same component, disjointness of the sets gives intrinsic separation
at least \(ch\), hence Euclidean separation at least \(c'h\) by
Lemma~\ref{lem:arrays-geometric-potential-preliminaries}(3).
Distinct components have a fixed positive distance, proving (2).

Finally, by (1), each \(P_a\) has measure at least \(c/N\) and lies in
\(B(z,R+C_1h)\) whenever \(x_a\in B(z,R)\). Since the \(P_a\) form a
partition, Lemma~\ref{lem:arrays-geometric-potential-preliminaries}(1)
and \(\nu(S)=1\) give
\[
 \frac cN\,\#\bigl(\Theta_N\cap B(z,R)\bigr)
 \leq \sum_{x_a\in B(z,R)}\nu(P_a)
 \leq \nu\bigl(B(z,R+C_1h)\bigr)
 \leq C(R+h)^2
 \leq C(R^2+N^{-1}).
\]
This proves (3).
\end{proof}

The next lemma bounds the sums of inverse powers of distances used below.

\begin{lemma}\label{lem:arrays-annular-counting}
Let $c_*,C_*>0$, and let $X=\{x_1,\dots,x_N\}\subset\R^3$ be a
set of $N\geq2$ distinct points such that
\[
 |x_a-x_{a'}|\geq c_*N^{-1/2}\quad(a\neq a'),
 \qquad
 \#\bigl(X\cap B(z,R)\bigr)\leq C_*(1+NR^2)\quad(z\in\R^3,\ R>0).
\]
Then for each integer \(j\geq1\), there is a constant \(C_j\), depending only
on \(j,c_*,C_*\), such that for any \(z\in\R^3\) and any nearest point
\(x_{a_*}\in X\),
\[
 \sum\limits_{a\neq a_*}|z-x_a|^{-j}\leq
 \begin{cases}
 C_1N,&j=1,\\
 C_2N\log N,&j=2,\\
 C_jN^{j/2},&j\geq3.
 \end{cases}
\]
In particular, these bounds hold at each point after omitting that point,
and apply to the arrays of
Lemma~\ref{lem:arrays-componentwise-equal-mass-cells}.
\end{lemma}

\begin{proof}
Set \(d=c_*N^{-1/2}\). After omitting a nearest point, separation and the
triangle inequality give \(|z-x_a|\geq d/2\). Points with \(|z-x_a|>1\)
contribute at most \(N\) in total. For the remaining points, the counting
hypothesis bounds the contribution of each dyadic annulus by
\[
 \sum_{\substack{a\neq a_*\\2^{k-1}d\leq|z-x_a|<2^kd}}
 |z-x_a|^{-j}\leq C_jN(2^kd)^{2-j},\qquad k\geq0.
\]
Summing over \(k\geq0\) with \(2^{k-1}d\leq1\) gives \(O(N)\) for \(j=1\),
\(O(N\log N)\) for \(j=2\), and \(O(Nd^{2-j})=O(N^{j/2})\)
for \(j\geq3\), as required.
\end{proof}

\begin{lemma}\label{lem:arrays-cell-quadrature}
Let $\nu$, $P_a$, and $x_a$ be as in
Lemma~\ref{lem:arrays-componentwise-equal-mass-cells}, and put
\[
 U_N(x)=\frac1N\sum\limits_{a=1}^N|x-x_a|^{-1}.
\]
Then, for all sufficiently large $N$,
\[
 \sup_a\Bigl|\frac1N\sum\limits_{a'\neq a}|x_a-x_{a'}|^{-1}-U_\nu(x_a)\Bigr|
 \leq CN^{-1/2}\log N.
\]
Moreover, $U_N\to U_\nu$ in $C^r(K)$ for every $r\geq0$ and every compact
$K\Subset\R^3\setminus S$.
\end{lemma}

\begin{proof}
Put \(h=N^{-1/2}\), fix \(a\), and set
\[
 J_a=\{a':|x_{a'}-x_a|\leq2C_1h\},
\]
where \(C_1\) is from \eqref{eq:cell-euclid-diam}. Replace the weights
\(1/N\) by \(\nu(P_{a'})\), and decompose the resulting error into the
weight correction, the contributions from the sets \(P_{a'}\) with
\(a'\in J_a\) (including \(P_a\)), and those from the remaining sets:
\[
 \frac1N\sum\limits_{a'\neq a}|x_a-x_{a'}|^{-1}-U_\nu(x_a)=E_{\mathrm{wt}}+E_{\mathrm{near}}+E_{\mathrm{far}}.
\]
The estimate for \(\nu(P_{a'})\) in Lemma~\ref{lem:arrays-componentwise-equal-mass-cells}
and Lemma~\ref{lem:arrays-annular-counting} give
\[
 |E_{\mathrm{wt}}|
 \leq CN^{-2}\sum\limits_{a'\neq a}|x_a-x_{a'}|^{-1}
 \leq CN^{-1}.
\]
There are \(O(1)\) sets \(P_{a'}\) with \(a'\in J_a\), the chosen points \(x_{a'}\) are separated by at least
\(ch\), and their union lies in \(B(x_a,3C_1h)\). Hence
Lemma~\ref{lem:arrays-geometric-potential-preliminaries}(1) gives
\[
 |E_{\mathrm{near}}|
 \leq \frac{C}{Nh}
   +\int_{S\cap B(x_a,3C_1h)}\frac{d\nu(y)}{|x_a-y|}
 \leq Ch.
\]
On each set \(P_{a'}\) with \(a'\notin J_a\), the mean value theorem bounds the kernel error by
\(Ch|x_a-x_{a'}|^{-2}\). Using \(\nu(P_{a'})\leq C/N\) and
Lemma~\ref{lem:arrays-annular-counting}, we obtain
\[
 |E_{\mathrm{far}}|
 \leq\frac{Ch}{N}\sum\limits_{a'\neq a}|x_a-x_{a'}|^{-2}
 \leq Ch\log N.
\]
These bounds are uniform in \(a\) and prove the first assertion.

For \(K\Subset\R^3\setminus S\), the differentiated Newton kernels and
their first derivatives in \(y\) are uniformly bounded near \(K\times S\).
The same estimates for the diameters and weights of the sets \(P_a\) therefore give
\[
 \|U_N-U_\nu\|_{C^r(K)}
 \leq C_{r,K}\left(h\sum\limits_a\nu(P_a)
       +\sum\limits_a|N^{-1}-\nu(P_a)|\right)
 \leq C_{r,K}(N^{-1/2}+N^{-1})\longrightarrow0.
\]
\end{proof}

Choosing \(\Theta_N=\{x_{a,N}\}\) as in
Lemma~\ref{lem:arrays-componentwise-equal-mass-cells} gives the desired
point-array construction. All assertions in
Theorem~\ref{thm:arrays-normalized-arrays} follow
directly from Lemmas~\ref{lem:arrays-componentwise-equal-mass-cells},
\ref{lem:arrays-annular-counting}, and~\ref{lem:arrays-cell-quadrature}.
In particular, scaling the estimate of
Lemma~\ref{lem:arrays-cell-quadrature} gives, uniformly in \(a\),
\[
 m_{a,N}=\frac{\psi(x_{a,N})}{v}+O(N^{-1/2}\log N)\geq\rho/2
\]
for all sufficiently large \(N\).
\section{The deformation of the glued solution}\label{sec:deformation}
In this section, we deform the glued approximate solution to a genuine solution.
Throughout, we work with rescaled constants \(e=v=1\), using the operators and
gluing construction of Section~\ref{sec:gluing}.
We begin with a local gauge estimate.

\begin{lemma}\label{lem:validation-mb027-radial-coefficient-gauge}

Let \((A_N,\Phi_N)\) be smooth configurations on \(\R^3\) such
that for every \(k\ge0\)
\begin{equation}
 \sup_{N,x}\left(
 |\nabla_{A_N}^kF_{A_N}(x)|+|\nabla_{A_N}^{k+1}\Phi_N(x)|
 +|\Phi_N(x)|\right)<\infty.                                  \label{eq:core-8}
\end{equation}
Then, for every \(y\in\R^3\), there is a gauge on \(B_2(y)\) in which
\begin{equation}
 \|A_N\|_{C^k(B_1(y))}+\|\Phi_N\|_{C^k(B_1(y))}\le C_k        \label{eq:core-9}
\end{equation}
uniformly in \(N,y\).

\end{lemma}

\begin{proof}

Trivialize by parallel transport along radial segments from \(y\).  In this
radial gauge, the nonabelian radial identity is
\begin{equation}
 (A_N)_i(y+x)=\int_0^1t x^j\,
 (F_{A_N})_{ji}(y+tx)\,dt,                                   \label{eq:core-10}
\end{equation}
Hence \(\|A_N\|_{C^0(B_1)}\) is controlled by
the curvature bound on \(B_2\). Differentiating \eqref{eq:core-10} gives
the connection part of \eqref{eq:core-9}. The identities
\[
 \partial_i\Phi_N=\nabla_{A_N,i}\Phi_N-[A_{N,i},\Phi_N]
\]
and their iterates then prove the Higgs part.  All segments used for points
of \(B_1(y)\) lie in \(B_2(y)\), so the constants are uniform.

\end{proof}

We also record a local elliptic estimate.

\begin{lemma}\label{lem:validation-mb028-local-first-order-estimates}
Let
\[
 D_{\mathrm{flat}}(a,\eta)=(*da-d\eta,d^*a),\qquad
 D=D_{\mathrm{flat}}+M(x)
\]
on \(B_1=B_1(0)\subset\R^3\). For \(1<p<\infty\), \(0<r<R\leq1\), and integers
\(m\geq0\), every distributional solution \(Du=f\), with
\(u\in L^p(B_R)\) and \(f\in W^{m,p}(B_R)\), satisfies
\begin{equation}\label{eq:nonlinear-10}
 \|u\|_{W^{m+1,p}(B_r)}
 \leq C\bigl(\|f\|_{W^{m,p}(B_R)}+\|u\|_{L^p(B_R)}\bigr).
\end{equation}
Here \(C\) depends on \(p,m,r,R\) and bounds for \(M\) in \(C^m\);
for \(m=0\), an \(L^\infty\) bound suffices.
\end{lemma}

\begin{proof}
The symbol of \(D_{\mathrm{flat}}\) satisfies
\begin{equation}\label{eq:core-47}
 |\sigma(D_{\mathrm{flat}})(\xi)(a,\eta)|^2
 =|\xi|^2(|a|^2+|\eta|^2).
\end{equation}
The multipliers \(i\xi_j|\xi|^{-2}\sigma(D_{\mathrm{flat}})(\xi)^*\)
are smooth away from the origin and homogeneous of degree zero.
The Mihlin multiplier theorem therefore gives
\(\|\partial w\|_p\leq C_p\|D_{\mathrm{flat}}w\|_p\) for compactly
supported smooth \(w\). If \(\chi\) is supported in \(B_R\) and equals
one on \(B_r\), then
\[
 D_{\mathrm{flat}}(\chi u)
 =\chi f+[D_{\mathrm{flat}},\chi]u-\chi Mu.
\]
Applying the flat estimate after mollification and passing to the limit
proves \eqref{eq:nonlinear-10} for \(m=0\). Commuting derivatives through
the equation and inducting on smaller concentric balls proves the higher
orders, with constants controlled by the stated coefficient bounds.

\end{proof}

In the rescaled variables, use the point arrays
\(\lambda_N\Theta_N=\{p_{1,N},\ldots,p_{N,N}\}\), where
\(p_{a,N}=\lambda_Nx_{a,N}\) are the normalized centers of
Theorem~\ref{thm:arrays-normalized-arrays}. Recall
\[
 \phi_N^{\mathrm D}(z)=1-\sum\limits_{a=1}^N|z-p_{a,N}|^{-1},
 \qquad
 m_{a,N}=1-\sum\limits_{b\ne a}|p_{a,N}-p_{b,N}|^{-1}.
\]
For fixed positive data \((v,q_0,\nu,\psi)\), the theorem gives
\(r_*\leq m_{a,N}\leq1\) for some fixed \(r_*>0\) and all sufficiently
large \(N\). For a fixed \(\alpha>0\), set
\[
 L_N=\alpha\log N,\qquad
 \mathcal S_{a,N}:=\mathcal S_{L_N}^{1,m_{a,N}}(p_{a,N})
 =\{z:3L_N/m_{a,N}\leq|z-p_{a,N}|\leq6L_N/m_{a,N}\}.
\]
Theorem~\ref{thm:arrays-normalized-arrays} gives
\(S_2(\lambda_N\Theta_N)\leq C(\log N)/N\) and separation at least \(c\sqrt N\).
Together with the mass bounds above, \(L_NS_2(\lambda_N\Theta_N)\to0\)
and \(L_N=o(\sqrt N)\) thus ensure the admissible mass--separation condition
for all sufficiently large \(N\).
We may therefore apply the gluing construction of
Section~\ref{sec:gluing} to these centers and masses, and denote the
resulting approximate solution by \(\Psi_N^G=(A_N^G,\Phi_N^G)\). Write
\[
 g_N:=(g_{\Psi_N^G},0)\in\Gamma(E),
\]
and let \(\widehat\sigma_N=\sigma_{\lambda_N\Theta_N}\) be the unit adjoint section
of the multi-center Dirac background.

By Proposition~\ref{prop:glued-approximate-basic-estimates}, the glued approximate solution
is smooth, has finite energy and charge \(N\), with \(e=v=1\) and
\[
 \{\Phi_N^G=0\}=\lambda_N\Theta_N,\qquad 0\leq|\Phi_N^G|\leq1,\qquad
 \operatorname{supp}g_N\subset\bigcup_{a=1}^N\mathcal S_{a,N}.
\]
Outside \(\bigcup_a B(p_{a,N},6L_N/m_{a,N})\), it agrees with the
multi-center Dirac background:
\[
 \Phi_N^G=\phi_N^{\mathrm D}\widehat\sigma_N,\qquad
 d_{A_N^G}\widehat\sigma_N=0,\qquad
 F_{A_N^G}=(*d\phi_N^{\mathrm D})\widehat\sigma_N.
\]
The following lemma gives the quantitative estimates needed for the deformation.

\begin{lemma}\label{lem:validation-mb027-many-core-approximate-pair}
For \(\alpha=\alpha(r_*)\) sufficiently large and all sufficiently large
\(N\), the glued approximate solution \(\Psi_N^G\) satisfies the following estimates.
\begin{enumerate}[label=\textup{(\arabic*)}]
\item Its residual satisfies
\[
 \|g_N\|_\infty\leq
 C\frac{\log N}{N},
\]
\[
 \|g_N\|_{6/5}=O\!\left(N^{-1/6}(\log N)^{7/2}\right),\qquad
 \|g_N\|_2=O\!\left(N^{-1/2}(\log N)^{5/2}\right).
\]
\item For each integer \(k\geq0\) and each \(y\in\R^3\), there is a gauge
over \(B_2(y)\) in which
\[
 \|A_N^G\|_{C^k(B_1(y))}+\|\Phi_N^G\|_{C^k(B_1(y))}\leq C_k,
\]
with \(C_k\) independent of \(N\) and \(y\).
\end{enumerate}
\end{lemma}

\begin{proof}
Suppress the index \(N\) on the points and masses, and put \(L=L_N\)
and \(S_2=S_2(\lambda_N\Theta_N)\).
Proposition~\ref{prop:glued-approximate-basic-estimates} gives
\begin{equation}\label{eq:core-26}
 \operatorname{supp}g_N\subset\bigcup_a\mathcal S_{a,N},\qquad
 \|g_N\|_\infty\leq C(S_2+e^{-cL}).
\end{equation}
Choose \(\alpha\) large enough that \(e^{-cL}\leq N^{-2}\), so
\(\|g_N\|_\infty\leq C(\log N)/N\).
The shells have total volume at most \(CNL^3\), so
\begin{equation}\label{eq:core-27}
 \|g_N\|_p\leq C\frac{\log N}{N}(NL^3)^{1/p}
 \qquad(p=6/5,\,2).
\end{equation}
Substituting \(L=\alpha\log N\) proves (1).

For (2), the explicit Prasad--Sommerfield formulas give uniform covariant derivative
bounds in the cores, since \(m_a\in[r_*,1]\).
Write \(A^D=A_{\lambda_N\Theta_N}\) and \(\nabla^D=\nabla_{A^D}\).
On the shells and exterior, the parallel Dirac section gives
\begin{equation}\label{eq:core-31}
 |(\nabla^D)^kF_{A^D}|
 +|(\nabla^D)^{k+1}(\phi_N^{\mathrm D}\widehat\sigma_N)|
 \leq C_k\sum\limits_a|x-p_a|^{-k-2}\leq C_k.
\end{equation}
Indeed, the nearest-point term is \(O(L^{-k-2})\), and the remaining
terms are bounded by Theorem~\ref{thm:arrays-normalized-arrays}.
Also \(0<\phi_N^{\mathrm D}\leq1\) there by
Lemma~\ref{lem:lower-mass-nonvanishing-on-shells} and the exterior
minimum principle.

On each shell, set \(a_a=A_N^G-A^D\) and
\(\varphi_a=\Phi_N^G-\phi_N^{\mathrm D}\widehat\sigma_N\).
Lemma~\ref{lem:local-ps-dirac-comparison-on-shell} and the cutoff bounds give
\begin{equation}\label{eq:core-31b}
 |(\nabla^D)^ka_a|+|(\nabla^D)^k\varphi_a|
 \leq C_k\left(\frac{m_a}{L}\right)^k
 \left(\frac{L}{m_a}S_2+e^{-c_kL}\right)\leq C_k.
\end{equation}
The curvature expansion, the product rule, and
\(\nabla_{A_N^G}=\nabla^D+\operatorname{ad}(a_a)\) then give uniform
bounds for the glued covariant derivatives on each shell. Combining the core,
shell, and exterior estimates gives
\begin{equation}\label{eq:core-32}
 \sup_{N,x}\left(
 |\nabla_{A_N^G}^kF_{A_N^G}|+
 |\nabla_{A_N^G}^{k+1}\Phi_N^G|+|\Phi_N^G|
 \right)\leq C_k.
\end{equation}
Lemma~\ref{lem:validation-mb027-radial-coefficient-gauge} now proves (2).
\end{proof}

For the rest of this section, write \(\nabla_N=\nabla_{A_N^G}\)
and \(D_N=D_{\Psi_N^G}\), with \(e=1\).
We next construct a uniformly bounded right inverse for the linearized operator.

\begin{lemma}\label{lem:validation-mb027-uniform-linear-right-inverse}

Put
\[
 Y=L^{6/5}(E)\cap L^2(E),
 \qquad
 X=L^2(E)\cap L^6(E),
\]
with their sum norms. For all sufficiently large \(N\), there is a bounded
linear map \(P_N:Y\to X\) such that \(D_NP_Nw=w\) distributionally for
every \(w\in Y\), and
\[
 \|P_Nw\|_X\leq C\|w\|_Y,
\]
where \(C\) is independent of \(N\).

\end{lemma}

\begin{proof}
By Lemma~\ref{lem:validation-mb027-many-core-approximate-pair},
\eqref{eq:weitzenbock-residual-bound}, and the H\"older, Sobolev, and Kato inequalities,
\[
 R(\Psi_N^G)\leq C\|g_{\Psi_N^G}\|_{3/2}
 \leq C N^{-1/3}(\log N)^3\longrightarrow0.
\]
Hence Proposition~\ref{prop:linear-right-inverse-coercive} applies for large
\(N\) and gives a right inverse \(P_N=P_{\Psi_N^G}\) with
\[
 \|P_Nw\|_2\leq\sqrt2\,\|w\|_{6/5}.
\]

For \(h\in L^2(E)\) with \(D_Nh\in L^2(E)\) in the distributional sense,
Lemma~\ref{lem:validation-mb028-local-first-order-estimates} with \(p=2,m=0\)
gives local \(H^1\) regularity. Applying its estimate in the uniform gauges of
Lemma~\ref{lem:validation-mb027-many-core-approximate-pair} and summing over
balls with uniformly bounded overlap, Kato and Sobolev yield
\begin{equation}\label{eq:core-45}
 \|\nabla_Nh\|_2+\|h\|_6\leq C\bigl(\|D_Nh\|_2+\|h\|_2\bigr),
\end{equation}
with \(C\) independent of \(N\).
Taking \(h=P_Nw\) and using \(D_NP_Nw=w\), we obtain
\[
 \|P_Nw\|_6\leq C(\|w\|_2+\|P_Nw\|_2)
 \leq C(\|w\|_2+\|w\|_{6/5}).
\]
Together these prove the uniform \(Y\to X\) estimate.
\end{proof}

We now use this right inverse to correct the approximate configurations
to exact solutions.

For \(h_i=(a_i,\eta_i)\), define the symmetric bilinear quadratic term
\begin{equation}
 \mathcal Q(h_1,h_2)=
 \left(
 \frac12*(a_1\wedge a_2+a_2\wedge a_1)
 -\frac12([a_1,\eta_2]+[a_2,\eta_1]),
 0
 \right).                                             \label{eq:nonlinear-2}
\end{equation}
Thus
\[
 \mathcal Q(h,h)=(* (a\wedge a)-[a,\eta],0).
\]

For a correction \(h_N=(a_N,\eta_N)\), write
\[
 \Psi_N=(A_N,\Phi_N):=(A_N^G+a_N,\Phi_N^G+\eta_N).
\]
The Bogomolny equation for \(\Psi_N\), together with the gauge condition
\(d_{A_N^G}^*a_N-[\Phi_N^G,\eta_N]=0\), is equivalent to
\begin{equation}\label{eq:nonlinear-5}
 D_Nh_N+g_N+\mathcal Q(h_N,h_N)=0.
\end{equation}

\begin{lemma}\label{lem:validation-mb028-small-defect-correction}
For every sufficiently large \(N\), there is a smooth correction \(h_N\)
satisfying \eqref{eq:nonlinear-5} with the following properties.
\begin{enumerate}[label=\textup{(\arabic*)}]
\item As \(N\to\infty\),
\begin{equation}\label{eq:nonlinear-6}
 \|h_N\|_X\leq C\|g_N\|_Y\longrightarrow0,\qquad
 \|\nabla_Nh_N\|_2+\|h_N\|_\infty\longrightarrow0,
\end{equation}
where \(C\) is independent of \(N\).
\item For each fixed \(N\),
\begin{equation}\label{eq:nonlinear-8}
 h_N(x)\longrightarrow0\qquad\text{as }|x|\longrightarrow\infty.
\end{equation}
\item The corrected configuration \(\Psi_N\) is a smooth finite-energy \(SU(2)\)
Bogomolny monopole with \(e=v=1\) and topological charge \(N\), satisfying
\begin{equation}\label{eq:nonlinear-9}
 0\leq|\Phi_N|\leq1\qquad\text{on }\R^3.
\end{equation}
\end{enumerate}
\end{lemma}

\begin{proof}
H\"older's inequality and interpolation between \(L^2\) and \(L^6\) give
\begin{equation}\label{eq:nonlinear-14}
 \|\mathcal Q(h,k)\|_Y\leq C_Q\|h\|_X\|k\|_X.
\end{equation}
Let \(C_P\) be the uniform right-inverse bound from
Lemma~\ref{lem:validation-mb027-uniform-linear-right-inverse}. The smooth map
\[
 \mathcal H_N(h)=h+P_N\mathcal Q(h,h),\qquad \mathcal H_N:X\to X,
\]
satisfies \(\mathcal H_N(0)=0\), \(d\mathcal H_N(0)=I\), and
\[
 \|d\mathcal H_N(h)-I\|_{X\to X}\leq2C_PC_Q\|h\|_X.
\]
Choose \(r>0\), independently of \(N\), with \(2C_PC_Qr\leq1/2\).
The quantitative Banach inverse function theorem gives a local inverse of \(\mathcal H_N\)
on \(B_X(0,r/2)\) with Lipschitz constant at most \(2\).\footnote{%
Alternatively, a more direct approach is to equip \(P_N(Y)\) with the
graph norm \(\|h\|_X+\|D_Nh\|_Y\) and apply the Banach inverse function theorem
directly to \(F_N(h)=g_N+D_Nh+\mathcal Q(h,h)\), with values in \(Y\).
Our approach is essentially the same, but avoids introducing the graph norm
at the cost of (probably) being less intuitive.}
Since \(\|g_N\|_Y\to0\), for large \(N\) we obtain
\begin{equation}\label{eq:nonlinear-24}
 \mathcal H_N(h_N)=-P_Ng_N,\qquad
 \|h_N\|_X\leq2C_P\|g_N\|_Y\longrightarrow0.
\end{equation}
Applying \(D_N\) and using \(D_NP_N=I\) proves
\eqref{eq:nonlinear-5}.

The global estimate \eqref{eq:core-45} gives
\begin{equation}\label{eq:nonlinear-26}
 \|\nabla_Nh_N\|_2
 \leq C\bigl(\|g_N\|_2+\|h_N\|_X+\|h_N\|_X^2\bigr)\longrightarrow0.
\end{equation}
In the uniform local gauges of
Lemma~\ref{lem:validation-mb027-many-core-approximate-pair}, the \(p=3\)
estimate of Lemma~\ref{lem:validation-mb028-local-first-order-estimates}
also gives
\[
 \sup_y\|h_N\|_{W^{1,3}(B_{1/2}(y))}
 \leq C\bigl(\|g_N\|_\infty+\|h_N\|_3+\|h_N\|_6^2\bigr)
 \longrightarrow0.
\]
Using \(W^{1,3}\hookrightarrow L^8\) and then the \(p=4\) estimate yields
\[
 \sup_y\|h_N\|_{W^{1,4}(B_{1/4}(y))}\longrightarrow0.
\]
Morrey's embedding gives \(\|h_N\|_\infty\to0\).
For fixed \(N\), smoothness follows by elliptic bootstrapping, since
\(W^{1,4}\) is a local algebra and the coefficients and \(g_N\) are smooth.

For each fixed \(N\), the uniform local \(W^{1,4}\) bound and Morrey's
inequality imply that the gauge-invariant norm \(|h_N|\) is uniformly
continuous. Since \(h_N\in L^2\), it follows that \(|h_N|\to0\) at
infinity, proving \eqref{eq:nonlinear-8}. In particular, \(|\Phi_N|\to1\) there.

By interpolation, \(h_N\in L^2\cap L^4\).
The background has finite energy; together with
\(\nabla_Nh_N\in L^2\), \(h_N\in L^2\cap L^4\), and
\(|\Phi_N^G|\leq1\), the curvature and covariant-Higgs expansions give
finite energy for the corrected solution.
On sufficiently large spheres, \(\Phi_N^G+t\eta_N\) is nonzero for
\(0\leq t\leq1\), so the normalized Higgs fields are homotopic and the
charge remains \(N\).
Finally, the Bogomolny equation and Bianchi identity imply
\[
 \tfrac12\Delta|\Phi_N|^2=|d_{A_N}\Phi_N|^2\geq0.
\]
The maximum principle and \(|\Phi_N|\to1\) at infinity give \(|\Phi_N|\leq1\).
\end{proof}

\section{The magnetic-bag limit}\label{sec:positive-limit}

In this section, we prove Theorem~\ref{thm:main}.
We first establish the limit for strictly positive surface charge
and a Higgs function with a positive lower bound. We then regularize the
prescribed nonnegative surface charge and conclude by a diagonal argument.

The following auxiliary lemma separates longitudinal elliptic estimates
from transverse exponential decay. Relevant arguments have also appeared in
Taubes~\cite{Taubes2013}, but we include the proof for completeness and
to address the specific requirements of our setting.

Let \(\Psi^0=(A^0,\Phi^0)\) be a smooth exact reducible
Bogomolny solution on \(B_{2R}(z_0)\) with \(e=1\) and
\(\Phi^0=|\Phi^0|\widehat\sigma\), where \(R\geq1\) and
\[
 d_{A^0}\widehat\sigma=0,\qquad |\widehat\sigma|=1,\qquad
 0<m\leq|\Phi^0|\leq1.
\]
Assume that
\begin{equation}\label{eq:bulk-1}
 \bigl\|\nabla^j|\Phi^0|\bigr\|_{L^\infty(B_{2R}(z_0))}\leq C_jR^{-j}
 \qquad(j\geq1),
\end{equation}
with \(m,C_j\) independent of \(R\). Let \(\mathcal Q\) be the quadratic term
from \eqref{eq:nonlinear-2}.
For a smooth correction \(h=(a,\eta)\), consider
\begin{equation}\label{eq:bulk-2}
 D_{\Psi^0}h+\mathcal Q(h,h)=0,\qquad
 \|h\|_{L^\infty(B_{2R}(z_0))}\leq\epsilon.
\end{equation}
Write \(h=h_L+h_T\), projecting both components onto
\(\mathbb R\widehat\sigma\) and its orthogonal complement, respectively.
Derivatives of \(h_L\) act on its real coefficients relative to the
parallel section \(\widehat\sigma\).

\begin{lemma}\label{lem:validation-mb029-large-empty-ball-abelian-response}
Under the above assumptions, there are \(R_0\geq1\), \(\epsilon_0,c>0\),
and constants \(C'_j\), depending only on \(m\) and
finitely many constants in \eqref{eq:bulk-1}, such that for
\(R\geq R_0\) and \(0<\epsilon\leq\epsilon_0\),
\begin{equation}\label{eq:bulk-3}
 |\nabla^jh_L(z_0)|
 \leq C'_j\epsilon R^{-j}
      +C'_j\epsilon^2e^{-cR}
 \qquad(j\geq1),
\end{equation}
and
\begin{equation}\label{eq:bulk-4}
 |(\nabla_{A^0})^jh_T(z_0)|
 \leq C'_j\epsilon e^{-cR}
 \qquad(j\geq0).
\end{equation}
\end{lemma}

\begin{proof}
We first establish transverse coercivity. The parallel splitting and
the bracket relations in \(\su\) give
\begin{equation}\label{eq:bulk-12}
 D_Lh_L=-\mathcal Q(h_T,h_T),\qquad
 D_Th_T+2\mathcal Q(h_L,h_T)=0,
\end{equation}
where \(D_L\) and \(D_T\) are the restrictions of \(D_{\Psi^0}\) to the
longitudinal and transverse subbundles, respectively, and
\(D_L(\alpha,\beta)=(*d\alpha-d\beta,d^*\alpha)\).
On the transverse subbundle, the Weitzenb\"ock formula is
\[
 D_T^\dagger D_T
 =\nabla_{A^0}^*\nabla_{A^0}+|\Phi^0|^2+W_T,
 \qquad |W_T|_{\mathrm{op}}\leq C\bigl|d|\Phi^0|\bigr|,
\]
where \(W_T\) is a zeroth-order endomorphism.
Thus, for every compactly supported transverse section \(k\) of \(E\),
\begin{equation}\label{eq:bulk-21}
 \|D_Tk\|_2^2\geq\|\nabla_{A^0}k\|_2^2+
 \int\bigl(|\Phi^0|^2-C\bigl|d|\Phi^0|\bigr|\bigr)|k|^2\,d^3x.
\end{equation}
Since \(|\Phi^0|\geq m\) and \(\bigl|d|\Phi^0|\bigr|\leq C_1/R\),
the last coefficient is at least \(m^2/2\) for large \(R\).
Using \(|[D_T,\chi]h_T|=|d\chi||h_T|\) and
\(|2\mathcal Q(h_L,h_T)|\leq C\epsilon|h_T|\), we apply
\eqref{eq:bulk-21} to \(\chi h_T\) and absorb the nonlinear term for small
\(\epsilon\) to obtain
\begin{equation}\label{eq:bulk-26}
 \|\nabla_{A^0}(\chi h_T)\|_2^2+\frac{m^2}{4}\|\chi h_T\|_2^2
 \leq2\int|d\chi|^2|h_T|^2\,d^3x.
\end{equation}

To obtain transverse decay, set \(\mathcal E(t)=\int_{B_t(z_0)}|h_T|^2\,d^3x\).
Choosing \(\chi\) supported in \(B_t\), equal to one on \(B_{t-1}\),
with uniformly bounded derivative, \eqref{eq:bulk-26} gives
\(\mathcal E(t-1)\leq C(\mathcal E(t)-\mathcal E(t-1))\).
Thus \(\mathcal E(t-1)\leq\theta\mathcal E(t)\) for a fixed
\(\theta<1\). Iterating from \(7R/4\) to \(5R/4\) yields
\begin{equation}\label{eq:bulk-28}
 \|h_T\|_{L^2(B_{5R/4}(z_0))}\leq C\epsilon R^{3/2}e^{-cR}.
\end{equation}
By Lemmas~\ref{lem:validation-mb027-radial-coefficient-gauge}
and~\ref{lem:validation-mb028-local-first-order-estimates}, interior
bootstrapping gives \(|(\nabla_{A^0})^jh|\leq C_j\epsilon\) on
\(B_{3R/2}(z_0)\). The homogeneous transverse equation in
\eqref{eq:bulk-12} thus has uniformly bounded coefficients and derivatives,
so its local elliptic estimates and \eqref{eq:bulk-28} yield
\begin{equation}\label{eq:bulk-35}
 \sup_{B_R(z_0)}|(\nabla_{A^0})^jh_T|
 \leq C_j\|h_T\|_{L^2(B_{5R/4}(z_0))}
 \leq C_j\epsilon R^{3/2}e^{-cR}.
\end{equation}
Decreasing \(c>0\) to absorb \(R^{3/2}\) proves \eqref{eq:bulk-4}.

Finally, we estimate the longitudinal derivatives. The product rule gives
\begin{equation}\label{eq:bulk-36}
 \|\nabla^k\mathcal Q(h_T,h_T)\|_{L^\infty(B_R(z_0))}
 \leq C_k\epsilon^2e^{-2cR}.
\end{equation}
The scaled interior estimate for the constant-coefficient operator
\(D_L\), from Lemma~\ref{lem:validation-mb028-local-first-order-estimates}
and Sobolev embedding, is
\begin{equation}\label{eq:bulk-37}
 |\nabla^jw(z_0)|
 \leq C_jR^{-j}\|w\|_\infty
 +C_j\sum\limits_{k=0}^{j}R^{k+1-j}\|\nabla^kD_Lw\|_\infty,
 \qquad j\geq1,
\end{equation}
with norms on \(B_R(z_0)\). Taking \(w=h_L\), using
\eqref{eq:bulk-12}, \(\|h_L\|_\infty\leq\epsilon\), and
\eqref{eq:bulk-36}, proves \eqref{eq:bulk-3} since \(R^{k+1-j}e^{-2cR}\leq C e^{-cR}\) for \(k\leq j\) and \(R\geq1\).
\end{proof}

We now return to the original scale, before the \(\lambda_N\) rescaling.
Fix positive data \((v,q_0,\nu,\psi)\) satisfying the hypotheses of
Theorem~\ref{thm:arrays-normalized-arrays}, with \(v=v_n\).
Let \((\widetilde A_N,\widetilde\Phi_N)\) be the corresponding monopoles from
Lemma~\ref{lem:validation-mb028-small-defect-correction}, with \(e=1\)
and \(|\widetilde\Phi_N|\to1\) at infinity.
Using the same \(\lambda_N\), set
\begin{equation}\label{eq:physical-5}
 e_N:=\frac{4\pi N}{q_0}=\frac{\lambda_N}{v},\qquad
 \delta_{\lambda_N}(x)=\lambda_Nx.
\end{equation}
Define the fields on the original scale by
\begin{equation}\label{eq:physical-11}
 A_N:=e_N^{-1}\delta_{\lambda_N}^*\widetilde A_N,
 \qquad
 \Phi_N:=v\delta_{\lambda_N}^*\widetilde\Phi_N,
\end{equation}
where the connection is pulled back as a one-form.

\begin{proposition}\label{prop:validation-mb029-fixed-regularization-physical-limit}
For all sufficiently large \(N\), the scaled configuration \((A_N,\Phi_N)\) is a
smooth finite-energy \(SU(2)\) Bogomolny monopole of coupling \(e_N\),
asymptotic mass \(v\), and topological charge \(N\), with
\begin{equation}\label{eq:physical-15}
 0\leq|\Phi_N|\leq v\qquad\text{on }\R^3.
\end{equation}
For every integer \(r\geq0\) and every compact \(K\Subset\R^3\setminus S\),
\begin{equation}\label{eq:physical-13}
 |\Phi_N|\longrightarrow\psi\quad\text{in }C^r(K).
\end{equation}
Consequently,
\begin{equation}\label{eq:physical-14}
 *d|\Phi_N|\longrightarrow*d\psi
 \quad\text{in }C^\infty(K).
\end{equation}
\end{proposition}

\begin{proof}
The scaling \eqref{eq:physical-11} gives
\[
 F_{A_N,e_N}=e_N^{-1}\delta_{\lambda_N}^*F_{\widetilde A_N},\qquad
 d_{A_N,e_N}\Phi_N=v\delta_{\lambda_N}^*
 (d_{\widetilde A_N}\widetilde\Phi_N).
\]
Since \(\delta_\lambda^*(*_z\omega)=\lambda *_x\delta_\lambda^*\omega\) for one-forms
and \(\lambda_N=e_Nv\), the scaled configuration solves the Bogomolny equation.
The remaining assertions on smoothness, finite energy, mass, charge,
and \eqref{eq:physical-15} follow from
Lemma~\ref{lem:validation-mb028-small-defect-correction} by scaling.

Theorem~\ref{thm:arrays-normalized-arrays} gives
\begin{equation}\label{eq:physical-7}
 v\phi_N^{\mathrm D}(\lambda_N\,\cdot)\longrightarrow\psi
 \quad\text{in }C^r_{\mathrm{loc}}(\R^3\setminus S)
 \quad\text{for every }r\geq0.
\end{equation}
Let \(h_N=(a_N,\eta_N)\) be the correction supplied by
Lemma~\ref{lem:validation-mb028-small-defect-correction}, and set
\(\epsilon_N=\|h_N\|_\infty\to0\).
Fix \(K\Subset\R^3\setminus S\) and set
\(R_N=\lambda_N\operatorname{dist}(K,S)/16\).
For \(x\in K\), the balls \(B_{2R_N}(\lambda_Nx)\) avoid all gluing
regions for large \(N\), since \(R_N\asymp N\) while the gluing radii
are \(O(\log N)\). On these balls the background is an exact reducible
solution and \(\widetilde\Phi_N=\phi_N^{\mathrm D}\widehat\sigma_N+\eta_N\).
Moreover, \eqref{eq:physical-7} gives uniformly
\begin{equation}\label{eq:physical-20}
 0<\frac{\inf\psi}{2v}\leq\phi_N^{\mathrm D}\leq1,\qquad
 \|\nabla_z^j\phi_N^{\mathrm D}\|_\infty
 \leq C_jR_N^{-j}\quad(j\geq1).
\end{equation}
Since the background preserves the splitting and \(e^{-cR_N}\leq C_jR_N^{-j}\),
Lemma~\ref{lem:validation-mb029-large-empty-ball-abelian-response} gives
\[
 |(\nabla_{A_N^G})^j\eta_N(\lambda_Nx)|\leq C_{j,K}\epsilon_NR_N^{-j}
 \qquad (x\in K,\ j\geq0).
\]
By \eqref{eq:physical-20} and \(\|\eta_N\|_\infty\leq\epsilon_N\), the segment
\(\Phi_N^G+t\eta_N\), \(0\leq t\leq1\), stays uniformly away from zero on
these balls for large \(N\). Hence
\[
 |\widetilde\Phi_N|-\phi_N^{\mathrm D}
 =\int_0^1\left\langle
 \frac{\Phi_N^G+t\eta_N}{|\Phi_N^G+t\eta_N|},\eta_N\right\rangle dt .
\]
By the product and chain rules, each differentiated term retains a factor
\(\epsilon_N\) from the last argument and gains \(R_N^{-1}\) per derivative.
Returning to the original scale multiplies each derivative by \(\lambda_N\).
Since \(R_N\asymp\lambda_N\), for every fixed \(r\) we obtain
\[
 \big\||\Phi_N|-v\phi_N^{\mathrm D}(\lambda_N\,\cdot)\big\|_{C^r(K)}
 \leq C_{r,K}\epsilon_N\longrightarrow0.
\]
Combining this with \eqref{eq:physical-7} proves
\eqref{eq:physical-13}; applying \(*d\) proves \eqref{eq:physical-14}.
\end{proof}

\paragraph{Proof of Theorem~\ref{thm:main}.}
We use the diagonal argument. Take \(v=v_n\) and the surface charge measure
\(\mu\) from \eqref{eq:density}, so that
\[
 q=\mu(S)>0,\qquad
 \phi=v-\frac1{4\pi}U_\mu\geq0\quad\text{on }\R^3.
\]
Choose a probability measure \(\nu_0\) with smooth, strictly positive
densities on every component of \(S\).
Choose \(t_s\downarrow0\) small enough and set
\begin{equation}\label{eq:arrays-positive-regularization}
 \mu_s=(1-t_s)\mu+t_s^2q\,\nu_0.
\end{equation}
Write
\[
 q_s=\mu_s(S),\qquad \nu_s=\mu_s/q_s,\qquad
 \phi_s=v-\frac1{4\pi}U_{\mu_s}.
\]
Then \(q_s=q(1-t_s+t_s^2)\in(0,q)\), and \(\nu_s\) has smooth, strictly
positive densities, bounded above and away from zero for each fixed \(s\).
Since \(\phi\geq0\) and \(U_{\nu_0}\) is bounded by
Lemma~\ref{lem:arrays-geometric-potential-preliminaries}(2), for \(t_s\) sufficiently small,
\begin{equation}\label{eq:arrays-positive-profile-bound}
 \phi_s=(1-t_s)\phi+t_sv-\frac{t_s^2q}{4\pi}U_{\nu_0}
 \geq t_s\left(v-\frac{t_sq}{4\pi}\|U_{\nu_0}\|_\infty\right)
 \geq t_sv/2>0.
\end{equation}
As \(s\to\infty\), the construction gives \(q_s\to q\) and
\(\mu_s\rightharpoonup\mu\); linearity
of the potentials and Lemma~\ref{lem:arrays-geometric-potential-preliminaries}(2)
give \(\phi_s\to\phi\) uniformly on \(\R^3\) and smoothly on compact sets
off \(S\).

Applying
Proposition~\ref{prop:validation-mb029-fixed-regularization-physical-limit}
with \((q_0,\nu,\psi)=(q_s,\nu_s,\phi_s)\) gives monopoles
\((A_{s,N},\Phi_{s,N})\) for each fixed \(s\) and all sufficiently large \(N\).
The compact sets \(E_k=\{x:|x|\leq k,\ \operatorname{dist}(x,S)\geq k^{-1}\}\)
exhaust \(\R^3\setminus S\). For each \(k\), first choose \(s_k\) by the
regularization convergence and then \(N_k\) by Proposition~\ref{prop:validation-mb029-fixed-regularization-physical-limit},
so that each approximation has error less than \(1/(2k)\) in \(C^{k+1}(E_k)\).
Choose \(N_k\) also large enough that
\(\bigl\||\Phi_{s_k,N_k}|-\phi_{s_k}\bigr\|_{C^0(E_k)}
<\frac12\inf_{\R^3}\phi_{s_k}\), ensuring that
\(|\Phi_{s_k,N_k}|\) is positive and smooth near \(E_k\).
Taking both sequences strictly increasing, the resulting monopoles
\((A^{(k)},\Phi^{(k)})=(A_{s_k,N_k},\Phi_{s_k,N_k})\) satisfy
\[
 \big\||\Phi^{(k)}|-\phi\big\|_{C^{k+1}(E_k)}<k^{-1},
\]
with norms on empty sets taken to be zero.

By Proposition~\ref{prop:validation-mb029-fixed-regularization-physical-limit},
these monopoles have finite energy, mass \(v\), charge \(N_k\),
coupling \(e_k=4\pi N_k/q_{s_k}\), and \(0\leq|\Phi^{(k)}|\leq v\).
Since \(q_{s_k}\to q>0\), we have \(e_k\to\infty\) and \eqref{eq:scaling}.
This estimate gives locally uniform convergence of the norms and smooth
convergence of their images under \(*d\) off \(S\). This proves
\textup{(i)} and \textup{(ii)}, since \(*d\phi=f_j\) on \(\Omega_j\),
with the projected-curvature interpretation on \(U_+\) explained in
Theorem~\ref{thm:main}.

Finally, since \(S\) has zero three-dimensional Lebesgue measure and is
hence negligible, the uniform bound and dominated convergence give
\(|\Phi^{(k)}|\to\phi\) in \(L^1_{\mathrm{loc}}\).
Applying \(*d\) distributionally proves \textup{(iii)}. \qed

\end{document}